**Polarized and Un-Polarized Emission from a Single Emitter in a Bullseye Resonator**

*Giora Peniakov\*, Quirin Buchinger, Mohamed Helal, Simon Betzold, Yorick Reum, Michele B. Rota, Giuseppe Ronco, Mattia Beccaceci, Tobias M. Krieger, Saimon F. Covre da Silva, Armando Rastelli, Rinaldo Trotta, Andreas Pfenning, Sven Höfling, and Tobias Huber-Loyola*

G. Peniakov, Q. Buchinger, M. Helal, S. Betzold, Y. Reum, A. Pfenning, S. Höfling, T. Huber-Loyola
Julius-Maximilians-Universität Würzburg, Physikalisches Institut, Lehrstuhl für Technische Physik, Am Hubland, 97074 Würzburg, Deutschland
E-mail: giora.peniakov@physik.uni-wuerzburg.de

M. B. Rota, G. Ronco, M. Beccaceci, R.Trotta
Department of Physics, Sapienza University of Rome, Piazzale Aldo Moro 5, 00185 Rome, Italy

T. M. Krieger, S. F. Covre da Silva, A. Rastelli
Institute of Semiconductor and Solid State Physics, Johannes Kepler University Linz, Altenberger Straße 69, 4040 Linz, Austria

Keywords: semiconductor quantum dots, bullseye microcavities, circular Bragg grating microcavities, polarization, single-photon emitters.

**Abstract**

We present polarized $|S| = 0.99 \pm 0.01$, and unpolarized $|S| = 0.03 \pm 0.01$ emission from a single emitter embedded in a single, cylindrically symmetric device design. We show that the polarization stems from a position offset of the single emitter with respect to the cavity center, which breaks the cylindrical symmetry, and a position-dependent coupling to the frequency degenerate eigenmodes of the resonator structure. The experimental results are interpreted by using numerical simulations and by experimental mapping of the polarization-resolved far-field emission patterns. Our findings can be generalized to any nanophotonic structure where two orthogonal eigenmodes are not fully spatially overlapping.

# 1. Introduction

Rapid progress in photonic quantum information technologies has highlighted the need for high-quality quantum light sources. As of today, quantum dots (QDs) in semiconductors provide the best functionality to fulfill this role.[1–3] They were demonstrated as a leading platform in producing not only single photons,[4–7] but also pairs and strings of entangled photons.[8–14] A way to optimize these capabilities is to use optical micro-cavities to support the QD emission. They can induce the Purcell effect on the QD emission, thus shortening the lifetimes of optical transitions, a beneficial trait for increasing coherence in entanglement generation schemes.[11–13] In addition, they typically support electromagnetic (EM) modes that allow efficient coupling of QD emission into optical fibers and waveguides.[15–17]

A leading design of optical micro-cavities currently explored is that of a circular Bragg reflector cavity, also known as a bullseye resonator.[18–21] It consists of a series of trenches, increasing in diameter that, due to high contrast in refractive index with the semiconductor environment, localize the EM field in their center. This design has drawn attention thanks to its relatively broad fundamental mode linewidth of the order of a few nanometers (in our wavelength range of interest, 800-950 nm), while maintaining a high Purcell effect of over 10 for a QD source residing in its center.[22,23] It was shown to support a Gaussian EM far-field mode and provide record-high light collection efficiencies.[16,17,22,23] These properties attracted attention not only in the QD community,[24–29] but also in studies of nitrogen-vacancy centers,[30] 2D materials,[31,32] polymers,[33] and gate-defined QDs in optical cavities.[34]

Depending on the targeted application, either polarized or un-polarized photon emission is required. For example, concepts and applications that exploit Hong-Ou-Mandel interference would benefit from polarized emission,[35] whereas unpolarized emission is essential for quantum entanglement schemes that use a one-to-one information translation between an anchored spin qubit and a photon [10–13] or polarization-entangled photon pairs.[8,22,23,36] Previously, it was suggested to fabricate elliptical cavities to break their symmetry and create two non-degenerate, orthogonal modes, which lead to polarized emission from the embedded source.[35,37,38] Another approach is to break the symmetry by fabricating grating with different periodicity in different direction.[39] Furthermore, high degrees of linear polarization were obtained from a linear dipole in a neutral QD embedded in unpolarized microcavities utilizing phonon-assisted excitation.[40]

In this work, we show that a displacement of the emitter from the center of the cavity can lead to strongly polarization dependent light-matter coupling and result in polarized emission, even from cavities that are themselves cylindrically symmetric and lead to completely unpolarized emission when the emitter is in the center of the cavity. We studied two samples with different material systems of QDs embedded in bullseye resonators. In both, we measured polarization-resolved photoluminescence (PL) from the resonator QDs and its cavity modes. In addition, we measured the polarization-resolved far-field emission of the resonators, containing the angle information of the polarized PL. Comparing these results to numerical simulations of the EM far-field, we claim that the QD emission profile is strongly governed by its displacement from the center of the bullseye resonator.

## 2. Studied Structures and the Experimental Setup

We studied two different self-assembled emitters in deterministically placed bullseye micro-resonators. We investigated two material systems in parallel: system 1 (system 2) consists of InGaAs (GaAs) QDs embedded in a GaAs ($Al_{0.33}Ga_{0.67}As$ surrounded by 4 nm GaAs on both sides) membrane with a thickness of $120 \pm 5$ nm ($140 \pm 5$ nm), which was grown using the Stranski-Krastanow technique followed by partial GaAs capping and annealing [24] (was grown by local droplet etching and nanohole filling [41]).

A $SiO_2$ ($Al_2O_3$) spacer layer of $360 \pm 10$ nm ($200 \pm 5$ nm) was sputtered (atomic layer deposited) on the membrane and later covered with gold to create a highly reflective mirror on the back side of the resonator. A flip-chip process was used to peel the GaAs ($Al_{0.33}Ga_{0.67}As$) membrane from the wafer substrate.[23,24] Using a hyperspectral (wide field) imaging technique,[42] we measured the position of the QDs relative to some gold (chromium) markers with a combined imaging and device fabrication precision of approximately 100 (50) nm. Then, we used electron-beam-lithography and dry etching to pattern bullseye resonator structures around the QDs. The central disk size of each bullseye resonator, its trench width and periodicity, were tailored to support an EM mode that matches the emission wavelength of the embedded QD (see **Figure 1**).

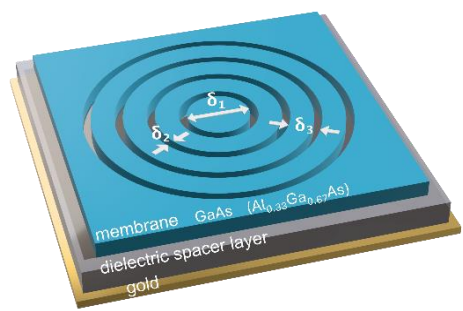

**Figure 1.** Bullseye resonator structure and dimensions for the GaAs (Al$_{0.33}$Ga$_{0.67}$As) membrane. Central disc diameter $\delta_1 = 754$ (666) nm. Trench width $\delta_2 = 100$ (100) nm. Trenches periodicity: $\delta_3 = 380$ (330) nm. Only four out of the seven (ten) trenches are shown.

To conduct spectroscopic studies on these structures, we used two similar confocal µPL setups. Both were operating at cryogenic conditions (~4 K) provided by closed-loop He cryocooler systems. We resolved the PL signal by a half-meter (0.75-meter) monochromator and recorded it using a charged coupled device camera. To analyze polarization, we used in the first setup a set of liquid crystal variable retarders allowing us to project the signal on the three Stokes parameters defined as

$$S_1 = \frac{I_H - I_V}{I_H + I_V}, \quad S_2 = \frac{I_D - I_A}{I_D + I_A}, \quad S_3 = \frac{I_R - I_L}{I_R + I_L}, \tag{1}$$

where $I_\alpha = |E_\alpha|^2$, $\alpha = H, V, D, A, R, L$ are the signal intensities for the horizontal, vertical, diagonal, anti-diagonal, right-hand, and left-hand polarization projections, respectively. In the second setup, the polarization was measured by a set of half-wave and quarter-wave plates.

**3. Polarization-Sensitive PL Measurements**

To quantify the polarization of the emission from a QD in a bullseye, we measured its PL projection on six different polarizations and calculated the integrated Stokes parameters over the whole emission wavelength range according to Equation 1. To exclude polarization effects from the setup, we introduced a polarization compensation by measuring the six prepared polarizations using a laser propagating though the optics. After this compensation, our Stokes coordinate system matched the lab frame. Measuring QDs outside the resonators gave an average polarization of $|S| = \sqrt{S_1^2 + S_2^2 + S_3^2} = 0.05 \pm 0.02$.

**Figure 2** presents the QD PL and cavity modes from two exemplarily chosen resonators. For the QDs emission in Figure 2 (a and c), we evaluated the total degree of polarization as its average in the range between 917.5-926 nm (793-796 nm), marked in the bottom panels of these figures by dashed vertical lines.

In Figure 2a, we present the unpolarized spectrum measured on the Al$_{0.33}$Ga$_{0.67}$As sample (system 2). Its integrated polarization was measured to be as low as $|S| = 0.04 \pm 0.03$. An even lower value of $|S| = 0.03 \pm 0.01$ was measured when restricting ourselves to the neutral exciton only (brightest feature at just below 794 nm). Both values are below the control reference of emission from QDs residing outside the resonators. To track the amount of possible splitting between the cavity modes, we show in Figure 2b their six polarization projections, normalized such that their maximal intensity is 1. No mode splitting was observed.

To estimate the errors in the reported values, we compared the contribution of three main sources: (i) Statistical noise of the measurement (detector noise) (ii) Our choice of wavelength integration range (dashed borders in Figures 2a and 2c, lower panels). (iii) The accuracy of our polarization-analyzing optics – either the liquid crystal variable retarders in the first setup, or the set of half- and quarter-wave plates in the second.

In the second sample, In(Ga)As QDs in a GaAs membrane (system 1), we measured strongly polarized PL (Figure 2c.). Most of it was measured horizontal, $S_1 = 0.98 \pm 0.01$, with a smaller anti-diagonal component of $S_2 = -0.19 \pm 0.01$. The circular component of $S_3 = 0.01 \pm 0.01$, was absent within the error. The total polarization-degree of this device was, therefore, $|S| = 0.99 \pm 0.01$. In this case, the modes spectrum revealed a minor splitting between the H- and V- modes of about 0.9 nm, probably arising from a small unintentional structural asymmetry of the resonator. Since the width of the modes and QD emission range spans over ~10 nm, approximately between 916-926 nm, and since their polarization degree is constant over this range, we dismiss mode splitting as the mechanism responsible for the polarized emission lines. Otherwise, we would expect changing polarization across this range – behavior we do not observe. Instead, we will argue in the rest of the paper that the polarization comes from the QD displacement relative to the center of the bullseye resonator. To support our claim, we will present in the next section an intuitive analysis of the EM cavity modes together with exact simulations.

Note that, the fact that the polarized and un-polarized emission was measured on different material systems is accidental - both polarization regimes can be achieved in both systems.

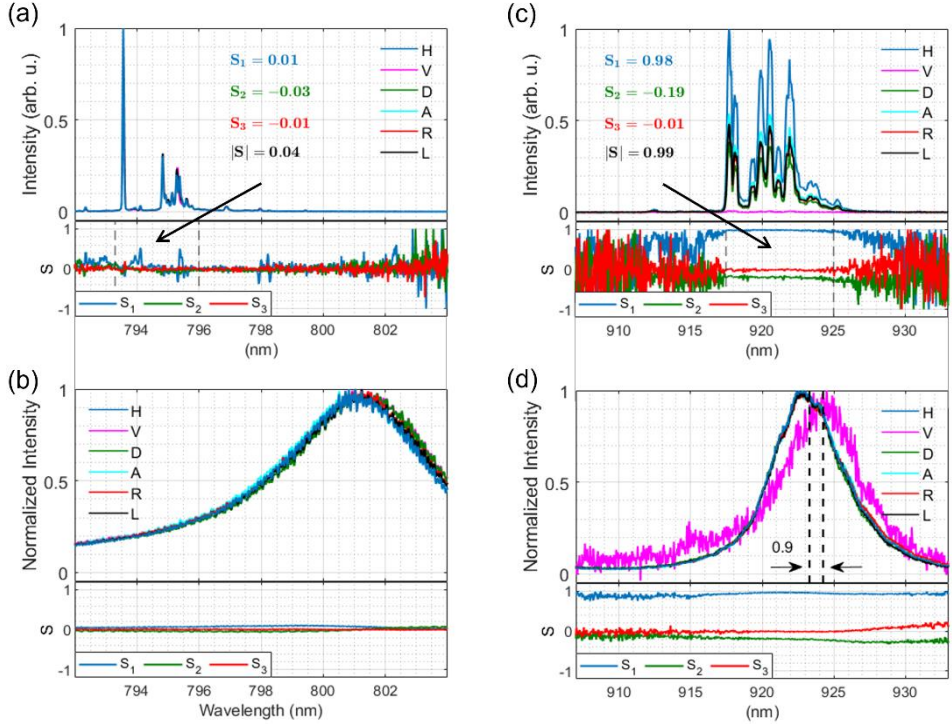

**Figure 2.** Polarization-sensitive PL of: (a-b) an un-polarized QD in a bullseye, (c-d) a polarized one. Figures a and c show the emission of the QDs at relatively low excitation power where single lines are resolved, whereas in Figures b and d the power was stronger to saturate the resonator modes. The bottom panel of each figure presents the polarization degree for different polarizations, given by the Stokes parameters as defined in Equation 1.

## 4. Displacement as Main Source of Polarized Emission

To theoretically investigate the source of polarized emission from the studied QDs in bullseye resonators, we performed finite-difference time domain simulations of the EM field within and outside the bullseye resonator using a commercial numerical solver. All the material and geometry parameters of our simulations were taken for the bullseye resonator in the GaAs membrane structure. We believe they capture the essential physics of similar resonator structures, including the second device we investigated in this work containing the $Al_{0.33}Ga_{0.67}As$ membrane. The QD is described as an unpolarized source by a model of a dipole with an orientation that is averaged over all in-plane directions, following Reference [43]. In this calculation, we define a dipole in the center of our coordinate system ($\vec{r} = 0$) with an orientation specified by $\varphi$, the in-plane azimuthal angle

$$\vec{P}(\varphi) = \vec{P}(0) \cos \varphi + \vec{P}\left(\frac{\pi}{2}\right) \sin \varphi \ . \tag{2}$$

This dipole generates an EM field that, due to the linearity of Maxwell equations, can be decomposed in a similar way

$$\vec{E}_\varphi(\vec{r}) = \vec{E}_{\varphi=0}(\vec{r}) \cos\varphi + \vec{E}_{\varphi=\pi/2}(\vec{r}) \sin\varphi \ . \tag{3}$$

Averaging the intensity of this field at point $\vec{r}$ over all dipole orientations, we find that it can be decomposed into a simple sum of only two field intensities, generated by two orthogonally oriented dipoles:

$$\langle |\vec{E}(\vec{r})|^2 \rangle = \frac{1}{2\pi} \int_0^{2\pi} |\vec{E}_\varphi(\vec{r})|^2 d\varphi = \frac{1}{2} \left( |\vec{E}_{\varphi=0}(\vec{r})|^2 + |\vec{E}_{\varphi=\pi/2}(\vec{r})|^2 \right) \ . \tag{4}$$

It is interesting to note that the EM field analysis is completely independent of the dipole coordinate system defining the angle $\varphi$. As long as the two dipoles are orthogonal ($\varphi = 0, \frac{\pi}{2}$), they generate the same field.

**4.1 Estimating the Degree-of-Polarization from the cavity EM Modes**

Calculating the EM field that a dipole pair generates when residing in a cylindrical symmetric resonator, one sums the contribution of all the cavity modes that has a non-vanishing field in its center. Performing the simulation for our particular bullseye resonator, we find that its circular distributed Bragg reflectors (DBRs), together with the longitudinal confinement in the Z direction (membrane thickness) support only one pair of these modes. This pair consists of two degenerate and orthogonal modes, each with three lobes within the central disk. The polarization of these modes is orthogonal, independent of the chosen basis. Here, we show that comparing the intensities of the two modes at any given point on the resonator plane well portrays the polarization that an emitter located at that point would emit. This can be understood considering the Purcell effect: depending on the local field intensity, the emitter couples with different strengths to the different modes, inducing polarized emission from the emitter.

To illustrate the above in an example, we show in **Figure 3a**, lower panel, the two supported orthogonal cavity modes in the H and V basis. Let us analyze the dependence of the emitter polarization on its displacement along the X direction. In the upper panels of Figure 3a, we present the cross-sections of the two modes in that direction. The emitted polarization for the displaced emitter is then expected to have a Stokes parameter $S_1$ according to the field intensities,

$$S_1 = \frac{|E_H|^2 - |E_V|^2}{|E_H|^2 + |E_V|^2} \ . \tag{5}$$

The other two Stokes parameters, $S_2$ and $S_3$ can be obtained by calculating the field intensities in the corresponding basis, $\{|E_D|^2, |E_A|^2\}$, and $\{|E_R|^2, |E_L|^2\}$, respectively.

The degree of polarization, captured by Equation 5, is presented in Figure 3b by the continuous blue curve. One can see that at zero displacement, $|E_H|^2$ and $|E_V|^2$ are equal, such that $S_1 = 0$. Moving along the X direction, the parallel polarization (H) is enhanced and peaks at 100 nm ($S_1 = 1$). Continuing to move along this direction, one observes first a decrease of $S_1$ down to negative values, corresponding to dominantly V-polarization, followed by an increase towards unpolarized emission and eventually a new decrease as the emitter approaches the edge of the central disk of the resonator. Consequently, several different values of displacement are possible for a given value of polarization.

It is important to note that our analysis omits the contribution of cavity modes that have a vanishing field at the resonator center. For a displaced emitter, these modes can, in general, be activated and contribute to the polarization profile. However, our comparison with the exact simulations, marked as orange crosses in Figure 3a and discussed in the next section, show that this contribution is negligible. Otherwise, one could add more dipoles in the resonator plane (not only in its center) to generate the more complete modes profile instead of those presented in Figure 3a.

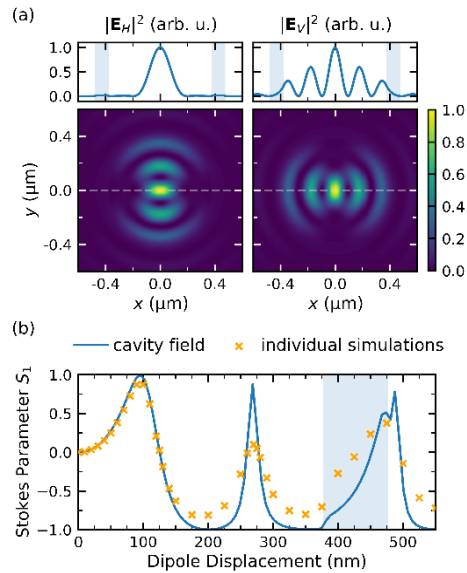

**Figure 3.** (a) Simulated bullseye resonator cavity modes in the H-V basis, presented in normalized intensity. The cross sections of these modes along the X axis are shown on the top part of each panel. (b) The degree of emitted linear polarization ($S_1$) as a function of the dipole pair displacement, for two calculation approaches. Blue curve: from cavity modes cross-section analysis (Equation 5); Orange crosses: net far-field polarization for individual

simulations per displaced dipole pair (see section 4.2). The light blue shadows in (a) and (b) mark the first trench of the bullseye structure.

## 4.2 Far Field Emission Simulations of a Displaced Emitter

To include the higher-order modes contribution in the resonator emission analysis, we repeated our finite-difference time domain simulations of the centrally placed dipole pair for discrete displacements along the X direction in steps of 10 nanometers. For each displacement, we calculated the polarization resolved far-field intensity one meter away from the resonator plane. An example of far field is shown in **Figure 4** for two displacements: (a) zero and (b) 80 nm displacement. In each simulation, we break the far-field profile into the three Stokes parameters $S_1, S_2, S_3$. Already for zero displacement, one sees that the linear Stokes parameters $S_1$ and $S_2$ are non-uniform across the entire region and show two-fold symmetric patterns. The directionality of these patterns follows the direction in which we chose to analyze the polarization. For example, in the $S_1$ plot, red and blue nodes appear along the X and Y direction, whereas in the $S_2$ plot, they appear along the diagonals. The value of $S_3$ is constantly zero over the whole region, in line with the cylindrical symmetry of the simulation setup.

This symmetry also accounts for the zero *net* values of the three Stokes parameters, defined as their integral over the entire far-field region. In Figure 4, we show the far-field patterns up to a conical angle of 54.1 degrees, matching the maximal collection angle of our objective in the experiment, having a numerical aperture of 0.81 (see section 5). Aside from the three Stokes parameter layouts, we present the total normalized emission intensity, showing its concentration in small emission angles up to 15° (second panel from the left).

For the 80 nm displaced dipole pair, we see a change in the Stokes parameter patterns compared to zero displacement. Generally, they lose their two-fold symmetry. Specifically, $S_1$ develops an oval-shaped pattern pointing in the direction of displacement, the symmetry of $S_2$ decreases, and $S_3$ develops a non-uniform pattern. In addition, the total emission gets slightly elongated in the X direction. Most importantly, the integrated (net) $S_1$ value seizes to be zero, becoming $S_1 = 0.74$. The net values of the other two Stokes parameters, $S_2$ and $S_3$, remain zero.

We performed a similar analysis for a set of displacements between 0 and 550 nm. We plot the net $S_1$ values for these displacements as orange crosses in Figure 3b, comparing them to the (blue) curve obtained by the previously discussed method (Section 4.1). Good agreement is observed. For the first ~150 nm, they are almost identical. Above that, some deviation appears, although both approaches keep agreeing on the positions of the minima and maxima points at $x = 270$ nm and $x = 480$ nm, for example. The deviation is the largest in the trench area (380-480 nm, blue-shaded area in the figure). However, in a real sample we cannot place an emitter in the trench since it consists of vacuum.

Evaluating the accuracy of the two approaches, we endorse the discrete displacement simulations approach (4.2) as the more accurate one. Unlike the cavity modes analysis (4.1), the discrete method does not neglect higher EM modes, and provides additional information such as the far field patterns. However, its accuracy comes with a computational cost: the cavity mode analysis requires substantially less computational overhead compared to the discrete method, solely because of the number of simulations involved. While in the first method, only one simulation of a dipole was needed, in the discrete displacement approach we conducted 73.

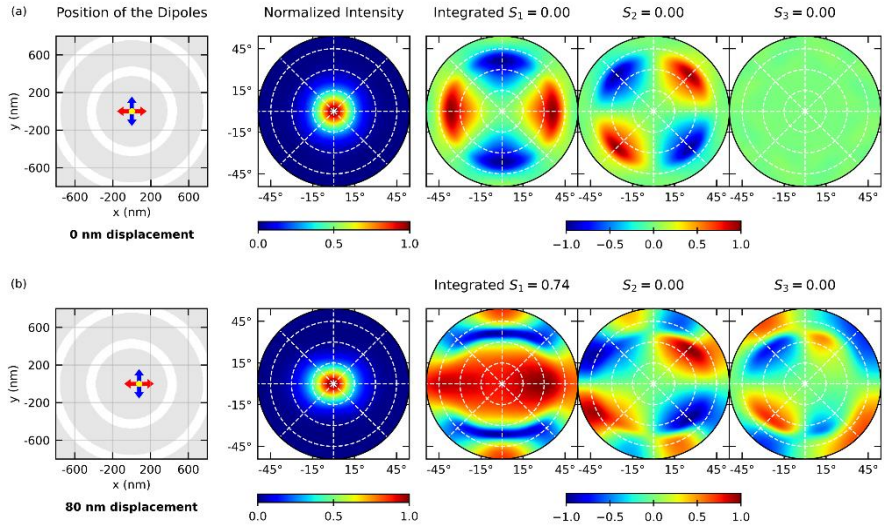

**Figure 4.** Far-field simulations of (a) a dipole pair placed in the center of a bullseye resonator, and (b) a dipole pair displaced by 80 nm in the X direction. The subplots, from left to right, are: (i) dipole pair position in the bullseye plane, shown up to the second trench (700 nm). (ii) total emission intensity, and (iii-iv) breakdown of the emission into the three Stokes components, $S_1, S_2, S_3$. The white dash circles mark emission angles of 15, 30, and 45 degrees. The net Stokes parameter, integrated over the whole emission angle range, is presented on top of each subplot.

## 5. Polarization-Sensitive Far-Field Patterns

For measuring the far-field emission of the bullseyes, we introduced a third lens to our μPL setup, positioned at one focal length distance from the objective's back focal plane. In this way, we mapped the collected signal angle onto position, which we later recorded on a camera. We projected the resulting image on the six polarization components, H, V, D, A, R, L, and calculated contrast images showing the angle profiles of the three Stokes parameters according to Equation 1. A set of these images for the polarized bullseye emission introduced in Figure 2c are shown in **Figure 5** next to the total emission.

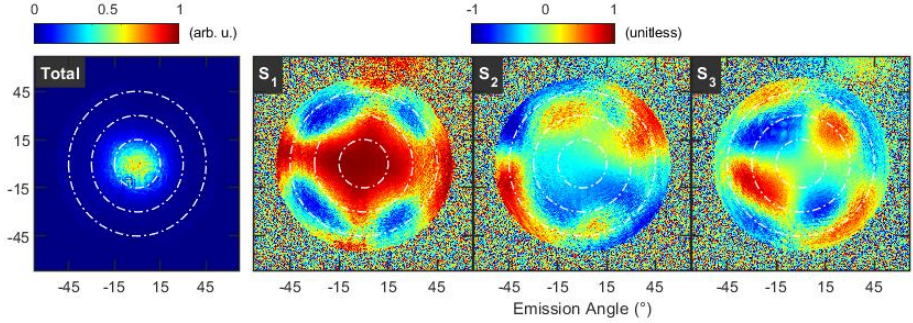

**Figure 5.** Polarization-sensitive far-field measurements of a QD in a bullseye. The most left subplot presents the total emission from the bullseye whereas the right three images are the breakdown of this emission into the three Stokes parameters $S_1, S_2, S_3$. Dash white circles mark emission angles of 15, 30, and 45 degrees for easy comparison to the simulations in Figure 4.

Comparing these images to the simulated far-field Stokes profiles for an 80 nm displaced dipole pair (Figure 4b), excellent agreement is observed. Many pattern features appear both in the measurements and simulations including the oval central spot along the X direction of $S_1$, and the alternating patterns appearing in $S_2$ and $S_3$. Integrating the three Stokes over the entire far-field circle, we obtain the net values $S_1 = 0.29, S_2 = -0.08, S_3 = -0.02$ (error up to 1 in the last digit). These values are lower than those reported for the same bullseye in Figure 2c, albeit the ratio between them is rather similar. This discrepancy stems from the fact that in the PL measurement, we analyzed polarization only in the wavelength emission range of the emitter, whereas in the far-field imaging, we collected emission from all wavelengths above 900 nm, cut by a long-pass filter. Therefore, in that case, the polarization of the QD is averaged out by the un-polarized background emission.

We conclude that our model of a displaced emitter reliably captures the primary mechanism responsible for polarizing the emission from the studied QD-embedded bullseye system.

## 6. Conclusion

We measured the emission polarization of QDs in bullseye resonators in two material systems and found examples of strongly polarized ($|S| = 0.99 \pm 0.01$) and un-polarized ($|S| = 0.03 \pm$ 0.01) devices. To study the origin of the polarization effect, we compared measurements and simulations of the resonators' far-field as projected on the three Stokes parameters, $S_1$, $S_2$, and $S_3$. In these simulations, we represented the QD emitter as a pair of two orthogonal dipoles and studied the effect of their displacement from the resonator center on the emitted polarization. We found that this effect can be neatly captured solely by considering their local coupling to the two degenerate and orthogonal cavity modes that are supported by the resonator. The strong resemblance between our measurements and simulations, together with the absence of a substantial measured mode splitting in these resonators, suggests that the QD displacement is the leading origin of polarization in those systems.

**Funding**

The German Ministry of Education and Research (BMBF) within the projects QR.X, Qecs and PhotonQ (Förderkennzeichen 16KISQ010, 13N16272 and 13N15759); the German Research Foundation (DFG) within the German Israeli Project Cooperation (DIP) grant "Bright sources of quantum light for efficient entanglement distribution"; the Austrian Science Fund (FWF) via the Research Group FG5, I 4320, I 4380, I 3762; the European Union's Horizon 2020 research and innovation program under Grant Agreements No. 899814 (Qurope), No. 871130 (Ascent+), and No. 101017733 (QD-E-QKD within the QuantERA II Programme); ICSC – Italian Research Center on High Performance Computing, Big Data and Quantum Computing, funded by European Union – NextGenerationEU.

**Acknowledgement**

The Linz Institute of Technology (LIT); the LIT Secure and Correct Systems Lab, supported by the State of Upper Austria; We thank Silke Kuhn for expert technical assistance.

**Conflict of Interest**

The authors declare no conflicts of interest.

**Data Availability Statement**

Data underlying the results presented in this paper are not publicly available at this time but may be obtained from the authors upon reasonable request.